\documentclass[
  ,draft            
  ,numberedheadings 
  ]
  {aipproc}

\usepackage[mathscr]{eucal}
 \usepackage{amsmath, amssymb}

  \def\sun{\hbox{$\odot$}}
   \def\degr{\hbox{$^\circ$}}
\newcommand{\be}{\begin{equation}}
\newcommand{\ee}{\end{equation}}
\newcommand{\bdm}{\begin{displaymath}}
\newcommand{\edm}{\end{displaymath}}
\def\la{\mathrel{\mathchoice {\vcenter{\offinterlineskip\halign{\hfil
$\displaystyle##$\hfil\cr<\cr\sim\cr}}}
{\vcenter{\offinterlineskip\halign{\hfil$\textstyle##$\hfil\cr
<\cr\sim\cr}}}
{\vcenter{\offinterlineskip\halign{\hfil$\scriptstyle##$\hfil\cr
<\cr\sim\cr}}}
{\vcenter{\offinterlineskip\halign{\hfil$\scriptscriptstyle##$\hfil\cr
<\cr\sim\cr}}}}}

\def\ga{\mathrel{\mathchoice {\vcenter{\offinterlineskip\halign{\hfil
$\displaystyle##$\hfil\cr>\cr\sim\cr}}}
{\vcenter{\offinterlineskip\halign{\hfil$\textstyle##$\hfil\cr
>\cr\sim\cr}}}
{\vcenter{\offinterlineskip\halign{\hfil$\scriptstyle##$\hfil\cr
>\cr\sim\cr}}}
{\vcenter{\offinterlineskip\halign{\hfil$\scriptscriptstyle##$\hfil\cr
>\cr\sim\cr}}}}}
\layoutstyle{6x9}

\begin{document}

\noindent {\it to appear in Astronomy Reports, July 2012}

\vspace{1cm}

\title{AE~Aquarii represents a new subclass of Cataclysmic Variables}

\classification{97.30.Qt}
\keywords{Accretion and accretion disks, magnetic field, binaries: close, white dwarfs, stars:
individual(AE~Aquarii)}

\author{N.R.\,Ikhsanov}{
  address={Pulkovo Observatory, Pulkovskoe Shosse 65, Saint-Petersburg 196140, Russia}
}

\author{N.G.\,Beskrovnaya}{
  address={Pulkovo Observatory, Pulkovskoe Shosse 65, Saint-Petersburg 196140, Russia}
}

\begin{abstract}
We analyze properties of the unique nova-like star AE~Aquarii identified with a close binary system containing a red dwarf and a very fast rotating magnetized white dwarf. It cannot be assigned to any of the three commonly adopted sub-classes of Cataclysmic Variables: Polars, Intermediate Polars, and Accreting non-magnetized White Dwarfs. Our study has shown that the white dwarf in AE~Aqr is in the ejector state and its dipole magnetic moment is $\mu \simeq 1.5 \times 10^{34}\,{\rm G\,cm^3}$. It switched into this state due to intensive mass exchange between the system components during a previous epoch. A high rate of disk accretion onto the white dwarf surface resulted in temporary screening of its magnetic field and spin-up of the white dwarf to its present spin period. Transition of the white dwarf to the ejector state had occurred at a final stage of the spin-up epoch as its magnetic field  emerged from the accreted plasma due to diffusion. In the frame of this scenario AE~Aqr represents a missing  link in the chain of Polars evolution and the white dwarf resembles a recycled pulsar.
\end{abstract}

\maketitle


  \section{Introduction}

 Cataclysmic Variables (CVs) are interacting low-mass close binaries containing a red dwarf (a normal component) and a white dwarf (a degenerate component). These systems emit variable radiation in a form of flares separated by phases of quiet (quasi-stationary) state. Depending on the amplitude, duration and recurrent time of the observed flares these objects can be classified as novae, dwarf novae and nova-like stars \cite{Warner-1995}. Their non-stationary behavior is caused mainly by mass exchange between the system components.  Matter lost by the red dwarf interacts with the degenerate component which lead to appearance of new sources of radiation responsible for peculiar features of CVs. Spatial and physical characteristics of these sources are determined by the parameters of the binary system and properties of its components. The stream of matter moving through the first Lagrangian point $L1$ to the white dwarf forms provided the red dwarf overfills its Roche lobe. The mass exchange in this case occurs on the dynamical time-scale \cite{Masevich-Tutukov-1988} and is characterized by a high ($\dot{M}_{\rm rd} \sim 10^{-9} - 10^{-7}\,{\rm M_{\sun}\,yr^{-1}}$) rate of mass-loss by the red dwarf. Evolution of a stream in the Roche lobe of the white dwarf depends on its spin period and magnetic field strength. On this basis CVs can be divided into three  main subclasses.

\noindent {\it \underline{I.~Non-magnetic}} Cataclysmic Variables are the systems in which the Alfv\'en  radius of the white dwarf, $R_{\rm A} = \kappa \left(\mu^2/\dot{M} (2GM_{\rm wd})^{1/2}\right)^{2/7}$, does not exceed its radius, $R_{\rm wd}$. Here $\mu$ and $M_{\rm wd}$ are the dipole magnetic moment and the mass of the white dwarf, $\dot{M}$ is the mass-transfer rate into its Roche lobe and  $\kappa$ is a parameter accounting for the geometry of the accreting flow ranging from 0.5 to 1 for the disk and spherical accretion, respectively  \cite{Ghosh-Lamb-1978}. Under this condition mass-transfer takes place via an accretion disk forming in the Roche lobe of the white dwarf and extending down to its surface.  Accretion of matter in the disk generates additional sources of radiation with total luminosity  $L_{\rm acc} = \dot{M} GM_{\rm wd}/R_{\rm wd}$. About one half of this energy is released at the inner radius of the disk in the region of its interaction with the surface of the white dwarf (a so called boundary layer)  and emitted in the ultraviolet (UV) and soft x-ray spectral domains. Accretion energy released in other parts of the disk is observed in the optical-UV part of the spectrum in a form of the continuum and  emission lines with typical double-peaked profiles (in the systems with high orbital inclination).

\noindent {\it \underline{II.~The Polars}} are CVs in which the spin period of the white dwarf is close to the orbital period of the system  and its magnetic field  is strong enough for its Alfv\'en  radius to exceed the radius of   circularization of the accreted material. The value of circularization radius for the case of the red dwarf filling its Roche lobe can be estimated as $R_{\rm circ} = a \left[(1+q)\left(0.5 - 0.227 \log{q}\right)^4\right]$, where $a$ is the orbital separation, $q = M_{\rm rd}/M_{\rm wd}$ is the mass ratio and $M_{\rm rd}$ is the mass of the red dwarf \cite{Frank-etal-2002}. Under these conditions the white dwarf magnetosphere prevents formation of a disk in its Roche lobe. At the same time the corotation radius of the white dwarf, $R_{\rm cor} = \left(GM_{\rm wd}/\omega_{\rm s}^2\right)^{1/3}$, turns out to exceed its Alfv\'en  radius due to a relatively slow (with a spin period of a few hours) axial rotation. Here $\omega_{\rm s} = 2 \pi/P_{\rm s}$ and $P_{\rm s}$ are the angular velocity and the spin period of the white dwarf. Centrifugal force at the magnetospheric boundary in this case does not prevent penetration of matter into the magnetic filed of the white dwarf and accretion onto its surface. Thus, a so called channeled accretion scenario can be realized in which the stream of matter flowing through the first Lagrangian point $L1$ enters the magnetosphere of the white dwarf without forming a disk, and flowing along the field lines reaches its surface in the magnetic pole regions. Additional source of radiation in this case is generated due to accretion and is localized at the base of the accretion channel near the stellar surface. The accretion luminosity is released predominantly in the X-ray and UV spectral range. A particular property of these objects (reflected in their name) is a high-degree circular polarization of their optical radiation. Understanding the cyclotron nature of this radiation and localization of its source near  the base of the accretion column made it possible to  establish independently that the surface magnetic fields of the white dwarfs in Polars  are in the range $20-100$\,MG \cite{Wickramasinghe-Ferrario-2000}.

\noindent {\it \underline{III.~Intermediate Polars}} contain white dwarfs with the Alfv\'en radius satisfying the condition  $R_{\rm wd} < R_{\rm A} < R_{\rm circ}$. Their spin periods span a wide range from tens of seconds up to an hour. However, all members of this subclass of Cataclysmic Variables are characterized by the corotation radius of a white dwarf being in excess of its Alfv\'en radius and hence the rotation of the white dwarf magnetosphere cannot hinder accretion of matter onto its surface. Under these conditions the infalling material can form a disk with the inner radius reaching the Alfv\'en radius of the white dwarf where the accreting flow penetrates into the magnetic field of the white dwarf and streaming along its field lines reaches the surface at the magnetic poles. The accretion energy released in the process of matter infall onto the surface of the white dwarf is radiated in a form of X-ray photons with an average energy in the range 6--10\,keV. In the case of oblique rotator (i.e. when the spin axis of the white dwarf is inclined to its magnetic axis) the X-ray source produces radiation pulsing at the spin frequency of the white dwarf. The disk emits optical-UV continuum with a number of emission lines often found to be double-peaked.

Thus, additional emitting source responsible for a peculiar appearance of Cataclysmic Variables arises due to accretion of matter lost by the red dwarf onto the surface of the white dwarf. Variability of these sources can be interpreted in terms of non-stationary character of the accretion process possibly connected with instabilities in the accretion flow and/or in the region of its interaction with magnetosphere of the white dwarf. High-amplitude flares are explained by  thermonuclear explosion in the matter being accumulated on the white dwarf surface in the process of accretion.

Classification outlined above is rather universal and spans practically all currently known Cataclysmic Variables with only a few exceptions the brightest of which is a low-mass close binary system AE~Aquarii. The next Section describes parameters of this system and its most interesting peculiarities. We will pay particular attention to  estimating the magnetic field of the white dwarf (Section 3) and show that the star is in the ejector state. Analyzing evolutionary status of the system in Section \,4 we conclude that spin and magnetic energy of the white dwarf had significantly increased during a previous epoch (no earlier than 10 million years ago) due to intensive accretion onto its surface. At the present time the system is in a particular state which we classify as ``Twister'' and discuss in Section \,5.

    \section{Main parameters and observational appearance of AE~Aquarii}

The beginning of AE~Aquarii study dates back to early 1930s, and throughout these 80 years the views of astronomical community on the nature of this system has changed dramatically over and over again. It was first described by Wachmann \cite{Wachmann-1931}, Zinner \cite{Zinner-1938} и Joy \cite{Joy-1943} as a nova-like star similar to  o~Ceti, T~Tauri and SS~Cygni, respectively. Henize \cite{Henize-1949} has noted that this star exhibits series of flares with a very short recurrence time close to one hour. AE~Aquarii
was the first Cataclsmic Variable found to be a spectroscopic binary containing a red dwarf with a hot companion \cite{Joy-1954}. The discovery of the 33s coherent oscillations in the optical \cite{Patterson-1979} and X-ray \cite{Patterson-etal-1980} emission of the system made it possible to   identify  a hot star with a fast rotating magnetized white dwarf. As a consequence AE~Aqr was assigned to the subclass of Intermediate Polars (DQ\,Her type stars) and until 1984 was considered a system wit non-stationary disk accretion onto a magnetized white dwarf. Observations of AE~Aqr with the Hubble Space Telescope \cite{Eracleous-etal-1994} allowed to estimate the average temperature  of the white dwarf atmosphere $T_{\rm wd} \sim 10\,000 - 16\,000$\,K and to identify a source of the 33\,s (and 16.5\,s) pulsations observed in the optical and UV spectral regions with two hot ($T_{\rm p} \sim 26\,000$\,K) spots located on opposite sides of the white dwarf and covering a surface of $S_{\rm p} \sim 10^{16}\,{\rm cm^2}$. Assuming that the spots are originated in the regions of magnetic poles the authors estimated inclination of the magnetic axis of the white dwarf to its spin axis in the range $76^{\degr} - 78^{\degr}$. But the most remarkable finding of the Hubble Space Telescope  observations was the fact that the intensity of pulsing component does not change during the flares thus arguing against the models associating flares with  accretion events.

Investigating evolution of the optical 33s-oscillations  during 14 years de~Jager et al. \cite{de-Jager-etal-1994} has made an intriguing discovery: the white dwarf is steadily spinning down at a rate  $\dot{P}_0 = 5.64\pm0.02 \times 10^{-14}\,{s\,s^{-1}}$. Further analysis by Welsh  \cite{Welsh-1999} confirmed a high stability of the white dwarf spin-down rate that ruled out a possibility to explain this result in terms of differential rotation of the white dwarf. The rapid breaking of the white dwarf implies the spin-down power
  \be
L_{\rm sd} = I \omega_{\rm s} \dot{\omega}_{\rm s} \simeq 6 \times 10^{33}\ I_{50}\ P_{33}^{-3}\ \left(\frac{\dot{P}}{\dot{P}_0}\right)\ {\rm erg\,s^{-1}},
  \ee
which exceeds the UV and X-ray luminosity of the system by a factor of 120--300 and even its bolometric luminosity, $L_{\rm bol}$, with account for the red dwarf radiation, by a factor of 5. Here $I_{50}$ and $P_{33}$  are the  moment of inertia and the spin period of the white dwarf in the units of $10^{50}\,{\rm g\,cm^2}$ and 33\,s. Thus, the spin-down power of the white dwarf dominates the energy budget of the system  ($L_{\rm sd} \gg L_{\rm bol}$) which according to the authors of this discovery is typical  for ejecting pulsars (i.e. neutron stars in the ejector state also known as radio-pulsars) and is unique not only for cataclysmic variables but also for a whole class of white dwarfs. This finding challenges commonly adopted assertion about an accretion nature of system radiation  in general.

Analyzing the observed Doppler H$\alpha$ tomogram the authors of \cite{Wynn-etal-1997} and  \cite{Welsh-etal-1998} concluded about an absence  of an accretion disk in the system. They have reported that the tomogram essentially varies from night to night, does not possess azimuthal symmetry and is not centered on the white dwarf. The main contribution into H$\alpha$ emission is made by the source whose velocity does not exceed $550\,{\rm km\,s^{-1}}$. Interpretation of this limitation in terms of Keplerian velocity, $v_{\rm k} = \left(GM_{\rm wd}/r\right)^{1/2}$, implies that the  emitting gas is situated at the distance $\geq 4 \times 10^{10}$\,cm from the white dwarf which exceeds its circularization radius by a factor of more than two (see Table\,\ref{t-3}). In addition, significant contribution is made by a source with spacial velocity less than $100\,{\rm km\,s^{-1}}$, that is two times less than the Keplerian velocity at the Roche surface of the white dwarf. The conclusion about the absence of an accretion disk in the system is also favored by single-peaked profiles of the emission lines in the optical and UV bands \cite{Reinsch-Beuermann-1994, Eracleous-etal-1994} as well as by a relatively small contribution of a source associated with matter moving inside the Roche lobe of the white dwarf to the optical radiation of the system \cite{van-Paradijs-etal-1989, Bruch-1991, Welsh-etal-1995}.

Data from X-ray observations also argues against an accretion nature of emission from AE~Aquarii. In some manifestations the X-ray spectrum of the object resembles coronal rather than accretion spectra \cite{de-Jager-1994}. It is much softer than those from Intermediate Polars: X-ray emission is dominated by photons in the energy range 0.1--1\,keV  \cite{Osborne-etal-1995, Choi-etal-1999, Choi-Dotani-2006}.
The X-ray luminosity of the system does not exceed that in the UV band and is less than luminosity of the emission source at the optical wavelengths. Contribution of the component pulsing at the spin period of the white dwarf does not exceed 18\% during the quiescence and decrease down to 7\% in flares. Analysis of the He-like $K\alpha$-triplet of nitrogen and oxygen  presented in \cite{Itoh-etal-2006} has shown that the linear size of the source of non-pulsing X-ray emission  exceeds the radius of the white dwarf by two orders of magnitude and the inferred plasma density of  $10^{11}\,{\rm cm^{-3}}$ is a few orders less than the estimated density in the accretion column of Cataclysmic Variables. This result was confirmed by further investigations \cite{Mauche-2009} which have revealed, however, that application of the same method to the line of Si~{\sc xiii} gives significantly larger values of plasma density.

In contrast to other Cataclysmic Variables AE~Aqr is a powerful source of non-thermal flaring radio-emission resembling some features of radio-emission from Cygnus\,X-3 \cite{Bastian-etal-1988}. The observed spectrum in a wide spectral range (from decimeter-wavelength radio to infrared, \cite{Abada-Simon-etal-2005}) exhibits a power-low ($\nu^{0.3-0.4}$) and can be described in terms of synchrotron mechanism. An absence of significant circular polarization implies that the observed radiation is generated by electrons with Lorenz factor  $\gamma \sim 0.3-30$ moving in the magnetic field with the strength ranging from 100 to 1000\,G. Flaring character of radiation is interpreted in the frame of van der Laan model \cite{van-der-Laan-1966} describing synchrotron radiation of electrons trapped in expanding plasmons \cite{Bastian-etal-1988}. There are some indications that the relativistic gas is ejected from the system at velocities close to 30\% of the speed of light \cite{de-Jager-1994, Abada-Simon-etal-2005}. The luminosity of the system in the radio band constitutes a small fraction of the spin-down power of the white dwarf: $L_{\rm r}/L_{\rm sd} \sim 10^{-5}$, that points out a significant contribution of the non-thermal processes to the total energy release in the system.

Some features of the ejector's appearance in the X-ray emission from AE~Aqr have been revealed in SUZAKU observations of the system. Analyzing the pulsing X-ray component in the energy range 10--30\,keV (first found in these observations), Terada et al. \cite{Terada-etal-2008, Terada-2010} made a conclusion about its non-thermal nature. They associated this radiation with relativistic electrons accelerated in the white dwarf magnetosphere. This was another argument in favor of ejection rather than accretion nature of emission from this source.

It is necessary to note that ejection activity of the source was somewhat exaggerated in the reports on observations of AE~Aqr in the high-energy ($\sim 10^{12}$\,eV) gamma-rays  (see, e.g., \cite{Meintjes-etal-1994, Chadwick-etal-1995}). However, these detections have not been confirmed by a detailed  analysis of 68.7\,hours of data recorded on AE~Aqr using the Whipple Observatory 10-meter gamma-ray telescope \cite{Lang-etal-1998}. Theoretical study of the processes of particle acceleration in the white dwarf magnetosphere \cite{Ikhsanov-Biermann-2006} has shown that this system is unlikely to be a source of intensive TeV emission. Finally, attempts to search for TeV gamma-ray emission from AE~Aqr with the MAGIC experiment were not successful either \cite{Sidro-etal-2008}. Thus, at present we have no grounds to believe AE~Aqr to be a Very High Energy gamma-ray source.


\begin{table}

\bigskip
\label{t-1}
\begin{tabular}{lccccc}
  \noalign{\smallskip}
  \hline
 {\it Parameters } & & & & & \\
 {\it System parameters} &  d,\,pc  & $P_{\rm orb}$,\,hr & $i$ & $e$ & $q$ \\
  \noalign{\smallskip}
  \hline
Value &  $(100\pm30)$  &  9.88 &  $50^{\degr}-70^{\degr}$ & $0.02$ & $0.6-0.8$ \\
  \hline
  \hline
  {\it Stellar } & & & & & \\
{\it Stellar parameters}   &  Type  & $M$ ($M_{\sun}$) & $P_{\rm s}$ &  $\dot{P}$ ($\mbox{s\,s$^{-1}$}$) & $\left[\vec{\Omega}\,\wedge\,\vec{m}\right]^{**}$\\
  \hline
Secondary  &  K3V--K5V   & $0.41\sin^{-3}{i}$ & 9.88\,hr & -- & --\\
Primary  & WD &  $0.54\sin^{-3}{i}$ & $33.08$\,s & $5.64\times 10^{-14}$ & $74^{\degr} - 76^{\degr}$ \\
 \noalign{\smallskip}
  \hline
    \noalign{\bigskip}
 \multicolumn{6}{l}{$^*$~~Detailed description of system parameters and corresponding references can be found in \cite{Ikhsanov-etal-2004b}} \\
 \multicolumn{6}{l}{$^{**}$~The angle between the spin and magnetic axes}\\
\end{tabular}
\caption{Parameters of AE~Aquarii$^{*}$}
\label{t-1}
\end{table}

The reliable system parameters established in the course of its continuous intensive study are collected in
Table\,\ref{t-1}. Limitations on the orbital inclination reflects that the system is non-eclipsing  ($i < 70^{\degr}$) and the mass of its degenerate component does not exceed Chandrasekhar limit ($i > 43^{\degr}$). The energy budget of the system is presented in Table\,\ref{t-2}. As already mentioned above, AE~Aqr is a non-thermal source in the radio and hard  (10--35\,keV) X-rays. The optical, UV and soft X-ray radiation of the system is predominantly thermal and is well fitted by superposition of three separate sources. The red dwarf dominates the visual light of the system (90--95\%, \cite{Bruch-1991}). Contribution of the white dwarf is seen  predominantly in the optical \cite{Patterson-1979}, UV \cite{Eracleous-etal-1994} and X-ray \cite{Choi-etal-1999} bands in the  form of pulsed emission modulated with its spin period. The third source is extended and highly variable. It manifests itself in the optical-UV continuum, broad single-peak emission lines as well as non-pulsing X-ray emission. This source is associated with the matter captured by the white dwarf from its companion and interacting with its rapidly spinning magnetosphere \cite{Eracleous-Horne-1996}. It is likely to be responsible for the unique flaring activity of the system: on the timescale from a few minutes to an hour the luminosity of the object in the blue pass-bands can change by an order of magnitude \cite{Beskrovnaya-etal-1996}. These flares show good correlation in the optical, UV and X-ray parts of the spectrum but do not correlate with radio flares which are remarkably similar in terms of amplitude and timescale \cite{Abada-Simone-etal-1995}.

The resemblance between AE~Aqr and other Intermediate Polars is limited to its undoubted membership in the class of Cataclysmic Variables (based on the nature of the binary components and non-stationary character of its radiation) together with a presence of coherent oscillations in the system emission. At the same time, this object dramatically  differs from all known Cataclysmic Variables with respect to both the nature of its energy source and a mechanism responsible for conversion of this energy to radiation emitted by the system. So unusual behavior of the system is presently associated with unique properties of its degenerate component whose parameters are mainly determined with enough accuracy. The only exception is the strength of its magnetic field which is a key parameter in modelling the processes of energy release and understanding the evolutionary stage of the system. In the next Section we will discuss the value of this parameter.

 \section{Magnetic field of the white dwarf}

Direct measurements of the white dwarf magnetic field by means of Zeeman spectroscopy is ineffective in case of AE~Aqr since photospheric lines of the primary cannot be identified in the spectrum of the system. On the other hand, estimation of its field strength by analogy with other Intermediate Polars is groundless because of radical difference between appearance of AE~Aqr and other objects of this subclass. In this situation it would be reasonable not to restrict oneself to similarity considerations but to explore a possibility to estimate this parameter using  indirect methods.

Discussion concerning the magnetic field of the degenerate component of AE~Aqr has started in the paper   \cite{Lamb-Patterson-1983}. The authors called attention to the fact that for the spin period of 33\,s the corotation radius of the white dwarf is only by a factor of 2--2.5 larger than its radius (see Table~\ref{t-3}). This means that even a relatively weak magnetic field on the surface of the white dwarf can prevent accretion of matter on its surface due centrifugal barrier at the magnetospheric boundary. That is why in the frame of popular at that time scenario of disk accretion the magnetic field strength of the white dwarf was believed to be below  $0.1$\,MG.


\begin{table}
   \begin{tabular}{ccc}
      \noalign{\smallskip}
        \hline
       \noalign{\smallskip}
Component  &  Quiescence  & Flares   \\
         \noalign{\smallskip}
        \hline
       \noalign{\smallskip}
Balmer continuum & $2.0 \times 10^{31}$ & $8.4 \times 10^{31}$ \\
        \noalign{\bigskip}
UV emission lines &  $1.6 \times 10^{31}$ & $4.1 \times 10^{31}$  \\
        \noalign{\bigskip}
H$\alpha$  & $4.8 \times 10^{30}$ & $1.4 \times 10^{31}$  \\
        \noalign{\bigskip}
X-rays, 0.1--5\,keV  & $7.8 \times 10^{30}$ & $1.7 \times 10^{31}$   \\
        \noalign{\bigskip}
Radio, 5--240\,MHz  & $10^{28}$ & $2 \times 10^{29}$   \\
        \noalign{\bigskip}
         \noalign{\smallskip}
        \hline
      \noalign{\bigskip}

 \multicolumn{3}{l}{Bolometric luminosity of the system ~~ $L_{\rm b}=10^{33}\,{\rm erg\,s^{-1}}$} \\
    \multicolumn{3}{l}{Spin-down power of the white dwarf ~~ $L_{\rm sd}= 6 \times 10^{33}\,{\rm erg\,s^{-1}}$} \\
        \end{tabular}
\caption{Energy budget of AE~Aquarii, erg\,s$^{-1}$}
\label{t-2}
    \end{table}

First doubts in reliability of this estimate were published by Bastian et al. \cite{Bastian-etal-1988}. Trying to explain the circular polarization of the optical radiation measured by Cropper \cite{Cropper-1986} within the accretion model (i.e. in terms of cyclotron radiation from the base of the accretion column, see \cite{Chanmugam-Frank-1987}), they came to conclusion that magnetic field of the white dwarf is in excess of 1\,MG. This result, however, leads to the following paradox: If the surface field of the white dwarf in AE~Aqr is indeed so strong, a steady accretion process onto its surface is impossible (its magnetospheric radius turns out to significantly exceed the corotation radius). But if there is no accretion on the surface of the white dwarf, then the source of polarized radiation suggested in paper \cite{Chanmugam-Frank-1987}, i.e. an accretion column, is absent. Thus, the result of Cropper (its reliability was later confirmed by Beskrovnaya et~al. \cite{Beskrovnaya-etal-1996}) proved to be the first argument in favor of revision of the magnetic field estimate and put under doubt accretion nature of radiation from this object.

The mass transfer modeling and interpretation of exceptionally rapid breaking of the white dwarf were the first steps in active revision of the existing picture of the system in general and of the magnetic field evaluation in particular. Results of these studies and analysis of possible origin of the circular polarization in the optical and  the pulsed component in hard X-rays are presented in this Section. We show that all available indirect methods for evaluation of the white dwarf magnetic field lead to the estimate of its dipole magnetic moment in the range $(1-2) \times 10^{34}\,\text{G\,cm}^3$.

   \subsection{Mass transfer}\label{mass-transfer}

Analyzing the energetic characteristics of the emission source one can conclude that the mass transfer rate in the system is in the range $10^{16} \la \dot{M} < 10^{18}\,{\rm g\,s^{-1}}$ \cite{Eracleous-Horne-1996, Ikhsanov-etal-2004a}. This means that the normal component is likely to overfill its Roche lobe and looses matter in a form of a stream through the first lagrangian point L1. But the matter flowing into the Roche lobe of the white dwarf neither accretes onto its surface nor accumulates around its magnetosphere forming a disk. In this situation the only assumption to make is that the matter is expelled from the system due to its interaction with the magnetic field of the rapidly spinning white dwarf. In other words, the primary in AE~Aqr acts as magnetic propeller.

\begin{figure}
\includegraphics{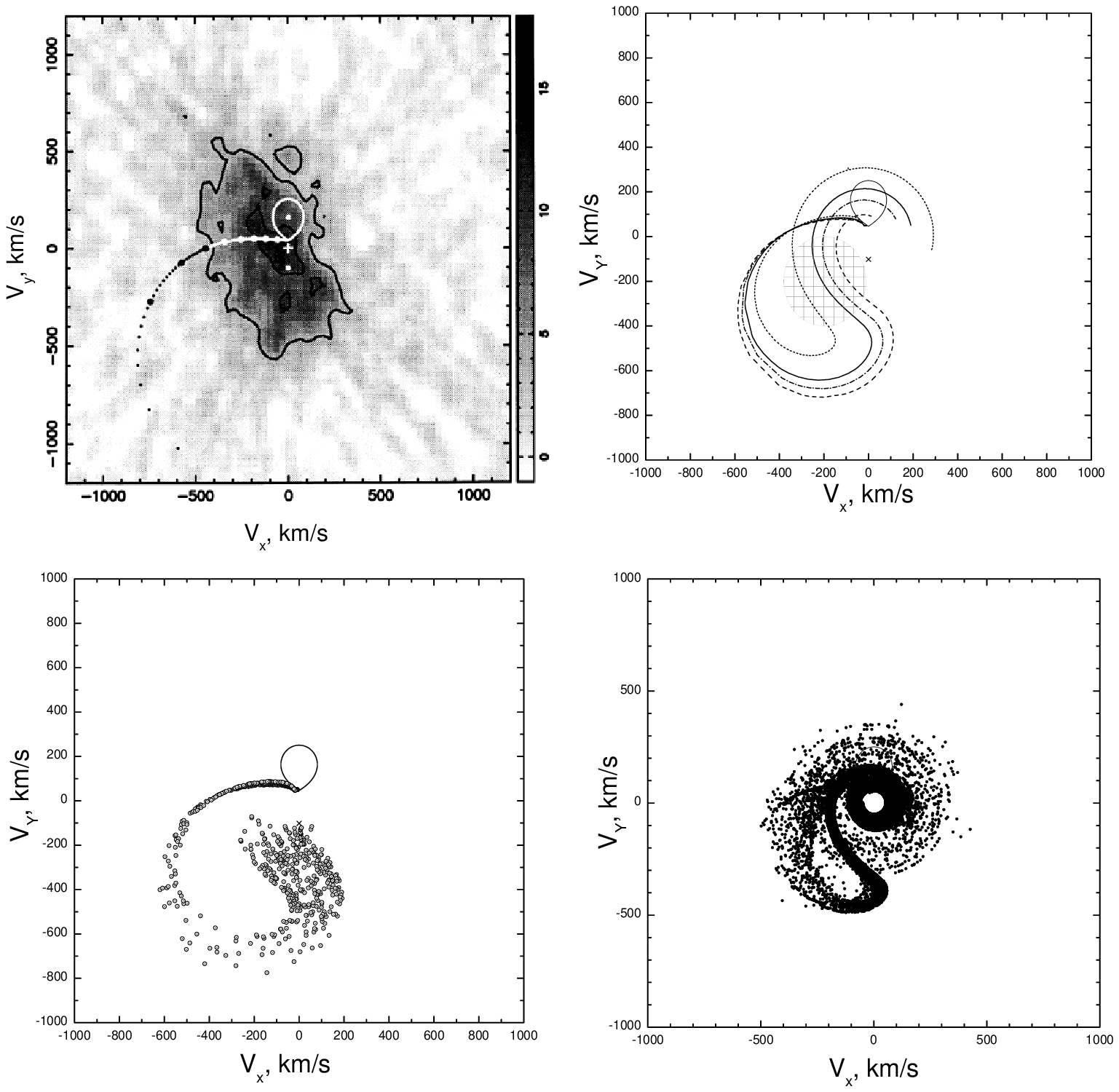}
 \caption{Doppler H$\alpha$ tomograms of AE~Aquarii. Left-up: observed
\cite{Welsh-etal-1998}; Left-down: simulated within magnetic propeller model by Wynn et al. \cite{Wynn-etal-1997}; Right-up: simulated  within magnetic propeller model
under the assumption that the emission source is situated in the region of blobs collisions \cite{Welsh-etal-1998}; Right-down: simulated within relativistic ejector model by Ikhsanov et al. \cite{Ikhsanov-etal-2004a, Ikhsanov-etal-2004b}}
\label{all-tomograms}
\end{figure}

Reliability of this assumption was first examined by Wynn et~al. \cite{Wynn-etal-1997}. Modelling interaction between rapidly rotating magnetosphere of the white dwarf and inhomogeneous  (fragmented into discrete diamagnetic blobs) stream via a surface drag term, they have concluded that ejection of  matter from  the system without forming a disk is possible only if  $\mu \ga 10^{32}\,\text{G\,cm}^3$. However, Doppler H$\alpha$  tomogram calculated under assumption that the dipole magnetic moment of the white dwarf is in the range typical for Intermediate Polars (i.e. $\mu \sim 10^{32}\,\text{G\,cm}^3$) contained high-velocity  ($|V| \sim 600-1000\,{\rm km\,s^{-1}}$) loop in the bottom-left quadrant which was not seen on the observed tomogram (see Fig.\,\ref{all-tomograms}).

%
\begin{table}
 \begin{tabular}{ccccccc}
    \noalign{\smallskip}
     \hline
      \noalign{\smallskip}
Radius   &  $R_{\rm wd}$ &  $R_{\rm cor}$ & $R_{\rm circ}$ & $R_{\rm L1}$ &  $R_{\rm lc}$ & $a$ \\
    \noalign{\smallskip}
     \hline
      \noalign{\bigskip}
value (cm) & ~ $6.5 \times 10^8$ & $1.5 \times 10^9$ &  $2 \times
10^{10}$  &~ $10^{11}$  ~ & $1.6 \times 10^{11}$ &  $(1.6-1.8) \times 10^{11}$ \\
      \noalign{\smallskip}
     \hline
      \noalign{\bigskip}
  \multicolumn{7}{l}{$R_{\rm wd}$ is the radius of the white dwarf;} \\
     \multicolumn{7}{l}{$R_{\rm cor}$ is the corotation radius;} \\
       \multicolumn{7}{l}{$R_{\rm circ}$ is the circularization radius;} \\
        \multicolumn{7}{l}{$R_{\rm L1}$ is the distance from the
white dwarf to the L1 point;} \\ 
   \multicolumn{7}{l}{$R_{\rm lc}$ is the radius of the light cylinder; $a$ is
the orbital separation of the system components.}\\
  \end{tabular}
\caption{Basic scales of AE~Aquarii.}
\label{t-3}
  \end{table}

Further elaboration of the tomogram within the same scenario but for different values of the white dwarf magnetic field \cite{Ikhsanov-etal-2004a, Ikhsanov-etal-2004b} have shown that the best agreement between the observed and simulated tomograms (avoiding the appearance of high-velocity loop) can be achieved for
$\mu \sim (1-2) \times 10^{34}\,\mbox{G\,cm$^3$}$. Under these conditions the stream approaches the white dwarf to a distance $R_0$ limited by the Alfv\'en radius of the white dwarf $R_0 \ga R_{\rm A}$. For typical parameters it can be expressed as
\be\label{r0}
R_{\rm A} \simeq 6 \times 10^{10}\ \eta_{0.37}\ \mu_{34.2}^{4/7}\ \dot{M}_{16}^{-2/7}\ M_{0.8}^{-1/7}\ {\rm cm},
 \ee
where $\mu_{34.2}=\mu/10^{34.2}\,{\rm G\,cm^3}$, $M_{0.8}$ is the mass of the white dwarf expressed in units of $0.8\,{\rm M_{\sun}}$,  $\dot{M}_{17}$, is the mass-transfer rate, expressed in units of $10^{17}\,{\rm g\,s^{-1}}$, and $\eta_{0.37}=\eta/0.37$ is the parameter accounting for the geometry of the accretion flow normalized following \cite{Hameury-etal-1986}. Under the conditions of interest $R_{\rm A}$ exceeds the
circularization radius (see Table\,\ref{t-3}) and, therefore, prevents a formation of an accretion disk in the system.

Alternative attempts to recalculate the tomogram presented by Wynn et al. \cite{Wynn-etal-1997} by making additional assumptions that in the process of interaction with the white dwarf magnetosphere the blobs remain cold \cite{Welsh-etal-1998} or on the contrary are heated to adiabatic ($\sim 5$\,keV) temperature \cite{Itoh-etal-2006} lead to conclusions contradicting results of observations. In particular, the maximum velocity on the tomogram for the case of cold blobs (shaded region in the Fig.\,\ref{all-tomograms}) is by a factor of 2 less than observed, while the mass transfer rate necessary to account for the system luminosity within this approach exceeds by an order of magnitude the upper limit on this parameter established from observations (see \cite{Ikhsanov-etal-2004a}). On the other hand, in the case of hot blobs the expected X-ray luminosity of the system  turns out to be by an order of magnitude higher than the observed luminosity and at least exceeds luminosity of the flaring source (see Table\,\ref{t-2}) which also contradicts observational data (see \cite{Ikhsanov-2006}).

Thus, study of mass-transfer picture in AE~Aqr by simulation of its Doppler H$\alpha$ tomogram suggests that the dipole magnetic moment of the white dwarf is $\mu \sim 1.5 \times 10^{34}\,{\rm G\,cm^3}$. The magnetic field strength on its surface  in the region of magnetic poles is
 \be\label{B0}
B_{\rm wd}^{\rm (p)} = \frac{2 \mu}{R_{\rm wd}^3} \simeq 100\,{\rm MG}\ \times R_{8.8}^{-3}\
\left[\frac{\mu}{1.5 \times 10^{34}\,{\rm G\,cm^3}}\right],
 \ee
and in the region of magnetic equator it is half this  value. Here $R_{8.8}$ is the white dwarf radius expressed in units of $10^{8.8}$\,cm.

   \subsection{Spin-down of the white dwarf}\label{spin-down}

Independent evaluation of the magnetic field strength of the white dwarf in AE~Aqr can be made investigating the mechanism responsible for its spin-down. The spin-down power of the white dwarf due to ejection of non-relativistic matter from the system (propeller action by the white dwarf) can be expressed as $L_{\rm p}^{\rm (k)} \sim \dot{M}_{\rm out} V_{\rm p}^2(R_0)$, where $\dot{M}_{\rm out}$ - is the mass ejection rate, and  $V_{\rm p}(R_0) = \left(2 GM_{\rm wd}/R_0\right)^{1/2}$ is the escape velocity of the matter at the distance of its closest approach to the white dwarf. It is necessary to take into account that the rate of release of the gravitational energy of the flow at the point of its closest approach to the white dwarf, $L_{\rm a}(R_0) \leq \dot{M}_{\rm out} GM_{\rm wd}/R_0$, cannot exceed  the bolometric luminosity of the extended source which according to observations in the optical, UV and X-ray bands is close to $10^{32}\,{\rm erg\,s^{-1}}$ (see Table\,\ref{t-2}). Combining these two conditions we get a simple expression $L_{\rm p}^{\rm (k)}(R_0) \sim 2 L_{\rm a}(R_0)$ which is a direct consequence of the virial theorem and does not depend on a mechanism by which the rotational energy of the compact object is transformed into the kinetic energy of the gas accelerated due to propeller action by the white dwarf. In other words, in the process of gas acceleration by the magnetic field of the white dwarf half of its spin-down power is transferred into the kinetic energy of the gas, and the second half - into its internal energy, e.g. heating. This conclusion is in accordance with the propeller model described in \cite{Shvartsman-1970, Lipunov-1980, Davies-Pringle-1981}.

An assumption that interaction between the stream which is a set of diamagnetic blobs and the magnetosphere takes place without any change of the internal energy of the interacting gas (i.e. the spin-down power of the white dwarf is completely converted into the kinetic energy of the matter ejected from the system, see e.g. \cite{Welsh-etal-1998, Pearson-etal-2003}) is incorrect since the interaction time of a stream moving along the ballistic trajectories within the Roche lobe of the white dwarf with its magnetic field does not exceed the free-fall time, \cite{Welsh-etal-1998, Pearson-etal-2003}, and hence is comparable to the time of release of the accretion energy of the flow. Furthermore, the rate of the spin-down power of the white dwarf conversion into the kinetic energy of the outflowing blobs under the conditions of interest is limited to \cite{Ikhsanov-Beskrovnaya-2008}
 \be\label{lpm}
L_{\rm p}^{\rm (m)}(r) \leq \frac{\mu^2}{2 \pi r^6} \Sigma_{\rm eff}(r)\ \dot{N}_{\rm b}(r)\ t_{\rm int}(r)\ |V_{\rm f}(r) - V_{\rm b}(r)|_{\bot},
 \ee
where $\Sigma_{\rm eff}$ is the effective cross-section of interaction between the magnetic field and the stream of blobs, $\dot{N}_{\rm b}$ is is the number of blobs interacting with the white dwarf magnetic field  in a unit time, $t_{\rm int}$  is the time of blob interaction with the magnetic field, and $|V_{\rm f} - V_{\rm b}|_{\bot}$  is the component of relative velocity between the stream of blobs and the magnetic field in the direction perpendicular to the field lines. The physical meaning of this expression is that the  kinetic energy of the blobs increases at the rate which is limited to the rate of magnetic flux transfer through the effective cross-section of interaction between the magnetic field and the blobs.

The number of blobs interacting with the white dwarf magnetic field  in a unit time at the distance of the stream closest approach to the white dwarf is limited as
 \be\label{dotnb}
 \dot{N}_{\rm b} \leq 3 \times \dot{M}_{17} \rho_{-11}^{-1} \ell_9^{-3},
 \ee
where $\rho_{-11}$ and $\ell_9$ are the density and radius of the blobs at the distance of closest approach in the units $10^{-11}\,{\rm g\,cm^{-3}}$ and $10^9$\,cm (normalized according to \cite{Wynn-etal-1997}).
Under the conditions of interest  ($R_0 \gg R_{\rm cor}$, see Table\,\ref{t-3}) the relative velocity between the blobs and the magnetic field can be approximated as $|V_{\rm f}(R_0) - V_{\rm b}(R_0)|_{\bot} \approx \omega_{\rm s} R_0$. The time of interaction between the blobs and the magnetosphere at the distance $R_0$ does not exceed the free-fall time, i.e. $t_{\rm int}(R_0) \leq t_{\rm ff}(R_0)$. Finally, the effective cross-section of interaction between the magnetic field and the blobs can be estimated as \be\label{sigma}
 \Sigma_{\rm eff} \simeq 2 \pi {\it l}_{\rm b} \delta_{\rm m} \sim 2 \pi {\ell}_{\rm b}
\left(D_{\rm eff}\ t_{\rm int}\right)^{1/2},
  \ee
where $\delta_{\rm m} = \left(D_{\rm eff} t_{\rm int}\right)^{1/2}$ is the thickness of diffusion layer at the surface of a blob determining the  scale of the magnetic field diffusion into a blob on a time scale   $t_{\rm int}$, and $D_{\rm eff}$ is the effective diffusion coefficient. Studies  of the solar wind penetration into the Earth magnetosphere  \cite{Gosling-etal-1991} have shown  that the maximum  value of $D_{\rm eff}$ is achieved in the case of Bohm diffusion ($D_{\rm B} = \alpha_{\rm B} \rho_{\rm i} c_{\rm s}$). Here $\rho_{\rm i}$ is the ion Larmor radius, $c_{\rm s}$ is the speed of sound in the region plasma interaction with magnetic field and   $\alpha$ is a coefficient accounting for non-elastic particle interaction and ranging within $0.1-0.25$ \cite{Gosling-etal-1991}.

Combining expressions ~(\ref{lpm}-\ref{sigma}), we find
 \begin{eqnarray}\label{lpm-final}
L_{\rm p}^{\rm (m)} & \la & 2 \times 10^{32}\,{\rm erg\,s^{-1}}\ \alpha_{0.25}\ \mu_{34.2}^{3/2}\ \dot{M}_{17}\ T_7^{1/2}\ \omega_{0.2}\ \ell_9^{-2}\ \rho_{-11}^{-1} \\
 \nonumber
 & & \times\ M_{0.8}^{-3/7}\ \left(\frac{r_0}{6 \times 10^{10}\,{\rm cm}}\right)^{-5/4} ~ \simeq 0.03\ L_{\rm sd},
  \end{eqnarray}
where $\alpha_{0.25}=\alpha/0.25$, $T_7$ is the plasma temperature in the region of interaction between the magnetic field and the blobs in units $10^7$\,K (normalized according to observed X-ray spectrum of the object) and  $\omega_{0.2} = \omega_{\rm s}/0.2\,{\rm rad\,s^{-1}}$. It is necessary to emphasize that the luminosity of the extended source in the frame of this scenario is in a good agreement with observational data and the description of interaction between the blobs and magnetic field does not contradict the virial theorem (see above). Thus, the maximum possible rate of the spin-down power of the white dwarf conversion into the kinetic energy of the gas due to propeller action by the white dwarf constitutes only a few percent of the observed spin-down power. This indicates that the observed breaking of the white dwarf is governed by a different mechanism.

As was quite correctly mentioned by the discoverers of the rapid breaking of the white dwarf in AE~Aqr \cite{de-Jager-1994, de-Jager-etal-1994}, the only objects of our Galaxy in which the spin-down power of the degenerate component significantly exceeds their bolometric luminosity  ($L_{\rm sd} \gg L_{\rm bol}$) are radio-pulsars (i.e. neutron stars in the ejector state)  (see, e.g., \cite{Lipunov-1987} and references therein). Usov \cite{Usov-1988} was the first to show that rapidly spinning white dwarfs can be in the ejector state provided their surface temperature satisfies the condition   $T_* \leq 10^6$\,K. In this case their spin-down power is released in a form low-frequency electromagnetic waves and relativistic wind and can be estimated with the use of expression for magneto-dipole losses, $L_{\rm md} = 2 \mu^2 \sin^2{\beta}\omega_{\rm s}^4/3 c^3$ \cite{Landau-Lifshits-1973}, where  $\beta$ is the angle between the magnetic and spin axes.

The surface temperature of the white dwarf in AE~Aqr estimated through observations with the Hubble Space telescope \cite{Eracleous-etal-1994} satisfies the condition from \cite{Usov-1988}. Therefore, there are no objections to application of the pulsar-like spin-down mechanism to description of the white dwarf breaking. Solving equation $L_{\rm sd} = L_{\rm md}$ for the dipole magnetic moment of the white dwarf we find that the observed spin-down rate can be explained within this scenario provided \cite{Ikhsanov-1998}
  \be\label{magmom}
\mu \simeq 1.4 \times 10^{34}\ P_{33}^{2} \left(\frac{L_{\rm sd}}{6 \times 10^{33}\,{\rm erg\,s^{-1}}}\right)^{1/2} {\rm G\,cm^3},
   \ee
where $P_{33}=P_{\rm s}/33$\,s, and the angle between the magnetic and spin axes of the white dwarf is adopted to be  $\beta = 77^{\degr}$ \cite{Eracleous-etal-1994}. This result is in good agreement with the estimate of the dipole magnetic moment obtained from simulation of the Doppler H$\alpha$ tomogram and implies that the strength of the dipole component of the surface magnetic filed of the white dwarf in AE~Aqr ranges from 50\,MG (at the magnetic equator) to 100\,MG (in the region of magnetic poles).

 \subsection{Circular polarization of the optical radiation}\label{circ-pol}

As already mentioned above the first attempts to explain the circular polarization of the optical radiation from AE~Aqr measured by Cropper  \cite{Cropper-1986}, led to a paradox thus indicating a necessity to reconsider estimates of the magnetic field strength of the primary component of this system. Reliability of these results have been confirmed by Beskrovnaya et al. \cite{Beskrovnaya-etal-1996} who have shown that the nightly-average value of circular polarization percentage in the $V+R$ passband is $0.06\%\pm0.01\%$ and is likely to vary on a timescale of the orbital period.

A polarized component has been reported to present in the optical radiation of Polars and some Intermediate Polars. Its origin is usually associated with cyclotron emission of plasma accreted onto the surface of the white dwarf in the region of its magnetic poles  \cite{Chanmugam-Frank-1987}. High-degree circular polarization of radiation from these systems is caused by significant contribution of accretion energy and, correspondingly, polarized component to the optical emission of these objects. Another point is that slow rotation of the white dwarfs in Polars (their spin periods are close to orbital periods of the binaries) makes it possible to integrate signal from only one pole during long exposure times (up to a few hours). The resulting degree of circular polarization in the optical measured from polars reaches extremely high levels up to tens of percents. The polarization percentage observed from rapidly rotating white dwarfs in Intermediate Polars  is significantly lower. The reason for this is weaker magnetic field of the primaries in Intermediate Polars and the fact that during the integration time both poles (producing polarization of opposite signs) are contributing to the resulting signal.

Application of this model to interpretation of polarization properties of AE~Aqr encounters, however, major difficulties. Observations with the Hubble Space Telescope \cite{Eracleous-etal-1994} have revealed that contribution of the hot spots on the surface of the white dwarf in the region of its magnetic poles to the system radiation in the $V+R$ passband does not exceed $0.1\%-0.2\%$. This means that the optical radiation coming from hot polar caps is diluted by a factor of $k_{\rm cap} \simeq 500-1000$. Besides, due to a very short spin period of the white dwarf in AE~Aqr, during an observing night we integrate contribution of both magnetic poles whose optical radiation is circularly polarized in the opposite directions.  The resulting degree of polarization depends on the system geometry. A detailed analysis of this situation by Ikhsanov et al.  \cite{Ikhsanov-etal-2002} has shown that due to effect of the white dwarf rotation the observed nightly-mean degree of circular polarization of AE~Aqr is by a factor of $k_{\rm pa} \simeq 4$ less than the intrinsic value. Summarizing, we come to conclusion that intrinsic circular polarization due to cyclotron radiation from the polar caps on the white dwarf surface should be $\ga 0.06\% \times k_{\rm cap} \times k_{\rm pa}$ that is in excess of 100\% in order to account for the observed value. This means that the hot spots on the white dwarf surface responsible for pulsed emission of the system in the UV/optical range cannot be the source of polarized radiation independent of their origin.

Optical circular polarization of white dwarf under certain conditions can be explained in terms of linear and quadratic Zeeman effect \cite{Jordan-1992}. High degree of polarization caused by this mechanism can be observed in some parts of the optical spectrum and is connected with broadening of photospheric lines in the strong magnetic field. The results of numerical simulations reported by Ikhsanov et al. \cite{Ikhsanov-etal-2002} show that radiation of the white dwarf in AE~Aqr should be polarized in the $V+R$ passband due to Zeeman effect provided the average value of magnetic field strength along its surface is in excess of 50\,MG. However, an expected value of polarization degree for magnetic field strength in the range 50--100\,MG is a factor of 4 smaller than the observed value. This discrepancy can be connected with  underestimate of the white dwarf magnetic field and/or its more complicated structure (e.g. significant contribution of multi-pole component) as well as with a presence of another source of polarized radiation. In the latter case measurements of circular polarization in the optical radiation from the system  cannot be used to evaluate magnetic field strength of its degenerate component. Thus, polarimetric data do not contradict estimates of the magnetic field obtained in Subsections 3.1 and 3.2.

 \subsection{Pulsing hard X-ray emission}\label{x-rays}

An independent estimate of the white dwarf magnetic field can be made on the basis of recently reported discovery of pulsed X-ray emission detected with the SUZAKU telescope in the energy range 10--30\,keV \cite{Terada-etal-2008}.  The authors of this discovery have associated this emission with radiative
losses of electrons accelerated in the magnetosphere of the white dwarf and hence has a non-thermal origin.  The luminosity of the system in this spectral range   has been evaluated as $\sim (0.5-2.3) \times 10^{30}\,{\rm erg\,s^{-1}}$ under assumption about isotropy of  the source. It has been noted, however, that the hard X-ray pulsations have a duty ratio of only 0.1, which allows to  conclude that the radiation is highly anisotropic with a beam angle not exceeding $40^{\degr}$ and to estimate its luminosity as \footnote{This value of $L_{\rm min}$ is smaller than that presented by \cite{Terada-etal-2008} by a factor of 100. It appears that the authors of \cite{Terada-etal-2008} have mistakenly evaluated the luminosity of the beamed source by multiplying the luminosity of the isotropic source by ($4 \pi/\gamma_{\rm col}$) instead of dividing it by the same value, where $\gamma_{\rm col}$ is the opening body angle of the beam.} $L_{\rm x-p} \ga L_{\rm min} \simeq 5 \times 10^{28}\,{\rm erg\,s^{-1}}$.

It is widely believed that acceleration of particles by the white dwarf in AE~Aqr is due to the fact that the electric potential in the magnetosphere of this star $V_{\rm e}^{\rm max} = (1/c) \omega_{\rm s} R_{\rm wd}^2 B_{\rm wd}$ can reach  huge values \footnote{$1$\,V = 300\,CGSE units}
\be
 V_{\rm e}^{\rm max} \sim 10^{17}\ B_8 P_{33}^{-1} R_{8.8}^2\ {\rm V},
 \ee
where $B_8 = B_{\rm wd}/10^8$\,G. This is not, however, a sufficient condition.   For particle acceleration  in this potential to be effective the number density of material in the region of acceleration should not exceed the Goldreich-Julian density
 \be\label{ngj}
n_{\rm GJ} = \frac{(\vec{\Omega} \cdot \vec{B})}{2 \pi c e} \simeq 5 \times 10^4\
\left(\frac{P_{\rm s}}{33\,{\rm s}}\right)^{-1}\ \left(\frac{B}{10^8\,{\rm G}}\right)\
{\rm cm^{-3}}.
 \ee

Otherwise, the electric field responsible for particle acceleration would be screened by the magnetospheric plasma. Furthermore, particle acceleration is caused by the electric field component parallel to the direction of the magnetic field lines (the electric field component perpendicular to the magnetic field lines cause only drift of particles (see, e.g. \cite{Artymovich-Sagdeev-1979}). As recently estimated by \cite{Ikhsanov-Biermann-2006} on the basis of the model by Arons and Scharlemann \cite{Arons-Scharlemann-1979}, this component is limited to $E_{\parallel} \leq E_{\rm AS}$, where
   \be\label{eas}
E_{\rm AS} = \frac{1}{8\sqrt{3}} \left(\frac{\omega_{\rm s} R_{\rm  wd}}{c}\right)^{5/2}
B(R_{\rm wd}).
 \ee

The electric potential responsible for particle acceleration in the magnetosphere of the white dwarf turns out to be
\be\label{phi}
 \varphi_{\rm as}(l_0) = \int_{\rm R_{\rm wd}}^{l_0} E_{\parallel}\ ds\ \simeq\
 2 \sqrt{2}\ E_{\rm AS}\ R_{\rm wd} \left[\left(\frac{l_0}{R_{\rm
wd}}\right)^{1/2} - 1\right]
 \ee
where $l_0=(R_{\rm wd}+s)$  is a distance from the surface of the white dwarf to the region of X-rays generation. The kinetic luminosity of the beam of relativistic particles accelerated in this potential on the spatial scale $r_{\rm wd} \ll l_0 \leq R_{\rm lc}$ is limited to  \cite{Ikhsanov-Biermann-2006}
  \be\label{dedt}
\dot{E}_{\rm p} \leq e \varphi_{\rm as}(l_0)\ \dot{N}\ \simeq 10^{32}\ B_8^2\ {\rm erg\,s^{-1}},
 \ee
where $e$ is the electron electric charge, and $R_{\rm lc} = c/\omega_{\rm s}$ is the light cylinder radius.
 \be\label{ndot}
 \dot{N} = \pi (\Delta  R_{\rm p})^2 n_{\rm GJ}(R_{\rm wd}) c,
 \ee
is the flux of relativistic particles from the polar caps of the white dwarf with the  radius
 \be\label{rp}
\Delta R_{\rm p} \simeq \left(\frac{\omega R_{\rm wd}}{c}\right)^{1/2} R_{\rm wd} \simeq 4 \times 10^7\ R_{8.8}^{3/2}\ P_{33}^{-1/2}\ {\rm cm}.
 \ee

As follows from Eq.~(\ref{dedt}), the observed luminosity of the hard X-ray pulsing component discovered by Terada et al. \cite{Terada-etal-2008} can be explained in terms of the pulsar-like acceleration mechanism only if the surface field of the white dwarf satisfies the condition
 \be
B_{\rm wd} \geq 30\ \eta_{0.1}^{-1} \left(\frac{L_{\rm hx}}{10^{29}\,{\rm erg\,s^{-1}}}\right)^{1/2} \left(\frac{\dot{\mathcal{E}}_{\rm p}}{10^{32}\,{\rm erg\,s^{-1}}}\right)^{-1/2}\ {\rm MG},
 \ee
where $\eta_{0.1}=\eta/0.1$ is the efficiency of conversion of the energy of accelerated particles into the energy of X-ray photons normalized to 10\%.

Thus, analysis of the key properties of AE~Aqr leads to the estimate of the  surface magnetic field of the white dwarf in this system in the range between 30 and 100\,MG. This is comparable to a typical magnetic field strength on the surface of the polars and is only an order of magnitude less than a record-breaking value of magnetic field measured in the white dwarfs \cite{Wickramasinghe-Ferrario-2000}. Following this result we can classify the state of the white dwarf in AE~Aqr as ejector. This reflects the fact that observed breaking of this star is  predominantly due to generation of the low-frequency waves and ejection of relativistic wind. Contribution of the propeller mechanism to the breaking of the white dwarf is insignificant.

 \section{Evolutionary state of the system}\label{evolution}

A conclusion about unusual state of the white dwarf in AE~Aqr rises questions on its origin and evolutionary state of the system in general. A hint to the answer is provided by a discrepancy between the age of the white dwarf determined by its cooling time and  the spin-down time scale  $t_{\rm sd} \simeq P_{\rm s}/2\dot{P} \simeq 10^7$\, years. Indeed, its age evaluated from the surface temperature  $T_{\rm wd} \sim 10\,000-16\,000$\,K and mass $M_{\rm wd} \sim (0.6-0.8)\,M_{\sun}$  is limited to $\ga 10^8$\,yr \cite{Schoenberner-etal-2000}, which exceeds by at least an order of magnitude the value of $t_{\rm sd}$. This implies that fast rotation of the white dwarf is not connected with peculiarities of its origin but is a product of the binary evolution which contained an epoch of rapid spin-up of the degenerate component  caused by intensive accretion onto its surface.

The spin period of the white dwarf can significantly decrease in the process of its accretion-driven spin-up only if $\dot{M}_{\rm pe} > \dot{M}_{\rm crit}$, where $\dot{M}_{\rm crit} \simeq 10^{-7}\,{\rm M_{\sun}\,yr^{-1}}$ is a critical value of the accretion rate at which the hydrogen burning in the matter deposited onto the white dwarf surface is stable  (see \cite{Livio-1995} and references therein). Otherwise, the spin behavior of the star will be similar to dwarf novae in which spin-up of the degenerate component is prevented by thermonuclear runaways on its surface  leading to nova outbursts followed by the expanded envelope mass-loss phase during which the accreted angular momentum is removed from the white dwarf \cite{Livio-Pringle-1998}.

Possible causes of the intensive mass-exchange stage in the history of AE~Aqr have been discussed by Meintjes \cite{Meintjes-2002}. In particular, he has pointed out that if the red dwarf overfills its Roche lobe, the mass transfer rate can reach $\dot{M}_{\rm pe} \sim 10^{19}-10^{20}\,{\rm g\,s^{-1}}$  ($\sim 10^{-7} - 10^{-6}\,{\rm M_{\sun}\,yr^{-1}}$) and keep this level during the whole spin-up epoch. If the magnetic field strength of the white dwarf, and, correspondingly, its magnetospheric radius, remains unchanged during this epoch, its spin period decrease down to 33\,s on the time scale  \cite{Ikhsanov-1999}
   \be\label{dur}
\Delta t_{\rm max} \geq \frac{2\pi I}{\dot{M}_{\rm pe}\ \sqrt{GM_{\rm wd} R_{\rm m}}}\
\left(\frac{1}{P} - \frac{1}{P_{\rm i}}\right) \simeq 2 \times 10^5\ I_{50} \dot{M}_{19}^{-1} M_{0.8}^{-1/2} R_9^{-1/2} P_{33}^{-1}\ {\rm yr},
    \ee
where $P_{\rm i}$ is an initial spin period of the white dwarf, which, as will be demonstrated below, satisfies the inequality $P_{\rm i} \gg 33$\,s.

The ultimate period which the white dwarf can reach in the process of disk accretion is given by
$P_{\rm min} = \max\{P_{\rm m}, P_{\rm eq}\}$, where $P_{\rm m}$ is a solution to equation  $R_{\rm s} = R_{\rm cor}$, and $P_{\rm eq}$ is an equilibrium period defined by equality of the spin-up torque, $K_{\rm su} = \dot{M}_{\rm pe} \left(GM_{\rm wd} R_{\rm s}\right)^{1/2}$, and spin-down torque, $K_{\rm sd} = (1/4) k_{\rm t} B_{\rm s}^2 R_{\rm wd}^6/R_{\rm cor}^3$, applied to the white dwarf from the accretion flow. Here $R_{\rm s}$ and $B_{\rm s}$ are the magnetospheric radius and the magnetic field strength on the surface of the white dwarf at the final stage of the accretion-driven spin-up, and  $k_{\rm t}$ is a numerical coefficient the most probable value of which is close to 0.3 (see \cite{Lipunov-1987}). In the case of stationary accretion and under the conditions of interest  $P_{\rm eq} \leq P_{\rm m}$. Taking $P_{\rm eq} = 33$\,s in equation  $K_{\rm su} = K_{\rm sd}$ and solving it for $B_{\rm s}$, we find that the observed spin period of the white dwarf in AE~Aqr can be reached within the scenario of accretion-induced spin-up provided  $B_{\rm s} \leq B_0$, where
 \be
 B_0 \simeq 1.5\,{\rm MG}\ k_{0.3}^{-7/12} M_{0.8}^{5/6} R_{8.8}^{-3} P_{33}^{7/6} \dot{M}_{19}^{1/2}.
 \ee
Here $k_{0.3} = k_{\rm t}/0.3$ and $\dot{M}_{19} = \dot{M}/10^{19}\,{\rm g\,s^{-1}}$. This means that
reconstructing the evolutionary track of the system it is necessary to take into account not only the spin evolution of the white dwarf (as has been done in \cite{Meintjes-2002}), but also the evolution of its magnetic field.

The magnetic field of the white dwarf may decrease during the spin-up epoch due to screening by the accreting material. \cite{Bisnovatyi-Kogan-Komberg-1974, Wickramasinghe-2006}. The hypothesis about a possibility to bury the magnetic field of accretors has been actively investigated for neutron stars \cite{Konar-Choudhuri-2004, Lovelace-etal-2005} and white dwarfs \cite{Cumming-2002}. The efficiency of screening has been shown to depend on the mass accretion rate and a duration of the intensive mass exchange between the system components. Under favorable conditions the surface magnetic field of a star can be reduced by a factor of 100. Afterwards the field is expected to reemerge in the process of diffusion through the layer of accreted plasma.

Following this hypothesis we can assume that prior to the epoch of active mass exchange the magnetic field strength on the surface of the white dwarf in AE~Aqr was close to its current value. At that time the system was likely to behave as a Polar (since the magnetospheric radius of the compact component under the condition  $\dot{M} \ll \dot{M}_{\rm pe}$ essentially exceeds its circularization radius). The start of spin-up epoch was caused by increase of the mass exchange rate up to $\dot{M}_{\rm pe} \geq 10^{19}\,{\rm g\,s^{-1}}$ due to red dwarf overfilling its Roche lobe. This resulted in decrease of the magnetospheric radius of the white dwarf down to $R_{\rm m}^{\rm (i)} \leq 10^{10}$\,cm and subsequent formation of the accretion disk in the system. The accretion of matter onto the surface of the white dwarf in this case could occur under condition $R_{\rm m}^{\rm (i)} < R_{\rm cor}$ which was satisfied provided the initial spin period of the white dwarf was $P_{\rm i} \geq 11$\,min.

The field of the compact object was found to be strongly screened by plasma accumulating in its polar caps  for accretion rates greater than the critical value $\geq 3 \times 10^{16}\,{\rm g\,s^{-1}}$ \cite{Cumming-2002}. Because of surface field decay the magnetospheric radius of the white dwarf is decreasing and, correspondingly, the area of the hot spots on its surface is increasing. The maximum possible factor of field reduction during the epoch of intensive accretion is limited to  $\sim \left(1/\sin{\theta_{\rm i}}\right)^{7/2} \sim 125$, where $\theta_{\rm i} = \arcsin\left(R_{\rm wd}/R_{\rm m}^{(i)}\right)^{1/2}$ is the opening angle of the accretion column at the beginning of spin-up epoch (see \cite{Bisnovatyi-Kogan-2006} and references therein). This implies that at the final stages of spin-up epoch the magnetic field of the white dwarf did not exceed 1\,MG and, hence, could not prevent decrease of the spin period down to its current value.

The spin-up time of the white dwarf  with account for screening of its magnetic field in the process of accretion can be evaluated by solving the equation $I \dot{\omega}_{\rm s} = \dot{M}_{\rm pe} \left(GM_{\rm wd} R_{\rm cor}\right)^{1/2}$ \cite{de-Jager-etal-1994} based on the assumption that the magnetospheric radius of the white dwarf is decreasing at the same rate that its corotation radius. The solution to this equation
  \be
\Delta t_{\rm min} = \frac{3}{4} \frac{(2 \pi)^{4/3} I}{\dot{M}_{\rm pe} (GM_{\rm wd})^{2/3} P_{\rm s}^{4/3}},
   \ee
determines the minimum duration of the spin-up epoch. The amount of matter accumulated on the white dwarf surface during this period can be estimated as
\be\label{deltama}
\Delta M_{\rm a} = \dot{M}_{\rm pe} \Delta t_{\rm min}  =  \frac{3}{4} \frac{(2 \pi)^{4/3} I}{(GM_{\rm wd})^{2/3} P_{\rm s}^{4/3}}.
  \ee
After the accretion epoch is over the surface magnetic field of the  white dwarf is gradually increasing due to diffusion of the buried field through the layer of screening plasma. The diffusion timescale of field can be estimated as $t_{\rm diff} \sim 4 \pi \sigma h^2/c^2$, where $\sigma$ is electron conductivity and $h = P/\rho g$ is pressure scale height. Here $P \sim 6.8 \times 10^{20} \rho_5^{5/3}\,{\rm erg\,cm^{-3}}$ is the pressure of non-relativistic degenerate gas, $\rho$ in the plasma density at the base of the screening layer  ($\rho_5 = \rho/10^5\,\text{g\,cm}^{-3}$) and $g = GM_{\rm wd}/R_{\rm wd}^2$. Using the value of electron conductivity of non-relativistic degenerate gas calculated in \cite{Yakovlev-Urpin-1980, Potekhin-etal-1995}, Cumming \cite{Cumming-2002} has shown that reemergence of the field of the white dwarf having undergone the stage of active accretion occurs on the timescale
 \be
\tau_{\rm diff} \simeq 3 \times 10^8 \left(\Delta M_{\rm a}/0.1\,{\rm M_{\sun}}\right)^{7/5}\,{\rm yr}.
 \ee

An appearance of a rapidly rotating  highly magnetic white dwarf can be expected only under the condition  $\tau_{\rm diff} \leq t_{\rm sd}$. Otherwise the spin period of the compact component will essentially increase on the timescale of field reemergence. Solving this inequality for the parameters of AE~Aqr we find
 \be
\Delta M_{\rm a} \leq 0.009\ P_{33}^{5/7} \left(\frac{\dot{P}}{5.64 \times 10^{-14}\,\text{s\,s}^{-1}}\right)^{-5/7}\ {\rm M_{\sun}}.
 \ee
Putting this value to Eq.~(\ref{deltama}) leads to a conclusion that the origin of an ejecting white dwarf in AE~Aqr can be explained in terms of accretion-induced spin-up provided its moment of inertia is
 \be
 I \leq 6 \times 10^{49}\ P_{33}^{4/3} \left(\frac{M_{\rm wd}}{{\rm M_{\sun}}}\right)^{2/3} \left(\frac{\Delta M_{\rm a}}{0.009\,{\rm M_{\sun}}}\right)\ {\rm g\,cm^2}.
 \ee
According to \cite{Andronov-Yavorskij-1990}, this condition is satisfied for white dwarfs with the mass in the range $1.1-1.2\,{\rm M_{\sun}}$.

The result obtained allows  to make some conclusions about the system parameters in general. First of all,
relatively large mass of the white dwarf indicates that the angle of orbital inclination is close to  $50^{\degr}$. This value is within the range of permitted values for this parameter \cite{Welsh-etal-1995}. It implies the mass of the red dwarf companion in excess of $0.7\,{\rm M_{\sun}}$ and, accounting for its tidal distortion  \cite{van-Paradijs-etal-1989}, lead to the estimate of its tidal radius (along the system major axis) comparable to the radius of its Roche lobe (see, also,  \cite{Ikhsanov-1997}). Besides, this result indicates that the spin-down power of the white dwarf can be overestimate by a factor of 2, and, accordingly, the dipole magnetic moment of the white dwarf is $\mu \simeq 10^{34}\,{\rm G\,cm^3}$ (see Eq.~\ref{magmom}). Finally, a correction of the inclination angle (its shift towards lower values) leads us to conclusion that the velocity of the gaseous stream in the Roche lobe of the white dwarf is somewhat greater than initially adopted and, hence, the distance of the stream closest approach  to the white dwarf is somewhat less than previously estimated. This fact has to be taken into account in the modeling of the mass transfer in the system in the present epoch.

 \section{Conclusions}\label{conclusion}

Our analysis has shown that most of the enigmatic properties of AE~Aqr result from  the presence of the white dwarf in the ejector state. Its origin is connected with intensive mass exchange between the system components  which started approximately 10 million years ago after a companion star had overflowed its Roche lobe. In the process of accretion which took place in that epoch, the material deposited from the accretion disk onto the white dwarf surface temporarily screened the internal magnetic field of the white dwarf thus making possible accretion-induced spin-up up to its current level. The transition of the white dwarf into the ejector state was caused by reemerging of the magnetic field by diffusion through the layer of accreted matter.

Relatively large age of the white dwarf  ($\sim 10^9$\,yr) derived from its average surface temperature, limitation on its intrinsic spin period ($P_{\rm i} > 11$\,min) and our estimate of its dipole magnetic moment ($\mu \sim 10^{34}\,\text{G\,cm}^3$) make us to suggest that before the spin-up epoch AE~Aqr could manifest itself as a Polar. During the spin-up epoch its X-ray luminosity exceeded $10^{36}\,{\rm erg\,s^{-1}}$ and the system could be seen as extremely bright Intermediate Polar. One cannot exclude that during the final phase of spin-up, the accretion of matter onto the white dwarf surface occurred directly from the accretion disk (as in non-magnetic CVs) and a  component pulsing at the spin period of the white dwarf was not present in the X-ray emission from the system. The duration of the present epoch is likely to be determined by the spin-down time-scale of the white dwarf which is close 10\,million years. At the end of this epoch one can expect dissipation of electric currents in the white dwarf magnetosphere and its transition to the propeller state. Further the system will appear as a Polar.

In the frame of this scenario AE~Aqr can be considered as a missing evolutionary link in the evolution of Polars, with its origin resembling in some aspects evolutionary scenario for recycled pulsars. At the same time, the analogy with  evolution of recycled pulsars is incomplete since before the spin-up epoch the white dwarf was in the accretor state with relatively slow rotation. Thus, in the case of AE~Aqr we deal with essentially new evolutionary stage of low-mass binaries requiring introduction of a new subclass which we call ``Twisters''. The degenerate objects in the systems from this subclass are in the ejector state. Intensive matter outflow from a system and a presence of high-luminous non-thermal component in its emission can be considered as indirect attribute of a Twister. Contribution of accretion luminosity to the energy budget of these systems is insignificant.

\begin{theacknowledgments}
The authors are grateful to G.S.\,Bisnovatyi-Kogan for critical discussions and useful comments. This work was supported by the Program of Presidium of Russian Academy of Sciences N\,21, and NSH-1625.2012.2.
\end{theacknowledgments}


\begin{thebibliography}{99}

\bibitem[Warner(1995)]{Warner-1995}
 B.\,Warner, \emph{Cataclysmic variable stars}, Cambridge: Cambridge Univ. Press (1995)

\bibitem[Masevich-Tutukov(1988)]{Masevich-Tutukov-1988}
A.G.\,Masevich, A.V.\,Tutukov, \emph{Stellar evolution: theory and observations} Moscow: Nauka (1988)

\bibitem[Ghosh-Lamb(1978)]{Ghosh-Lamb-1978}
P.\,Ghosh, F.K.\,Lamb, \emph{Astrophys. J.}, \textbf{223}, L83 (1978)


\bibitem[Wickramasinghe-Ferrario(2000)]{Wickramasinghe-Ferrario-2000}
 D.T.\,Wickramasinghe, L.\,Ferrario, \emph{PASP} \textbf{112}, 873 (2000)

\bibitem[Frank-etal(2002)]{Frank-etal-2002}
 J.F.\,Frank, A.R.\,King, D.J.\,Raine, \emph{Accretion power in Astrophysics}, Cambridge: Cambridge Univ. Press (2002)

\bibitem[Wachmann(1931)]{Wachmann-1931}
 A.A.\,Wachmann, \emph{Astonomische Nachrichten} \textbf{242}, 382 (1931)

\bibitem[Zinner(1938)]{Zinner-1938}
 E.\,Zinner, \emph{Astonomische Nachrichten} \textbf{265}, 345 (1938)

\bibitem[Joy(1943)]{Joy-1943}
 A.H.\,Joy, \emph{PASP} \textbf{55}, 283 (1943)

\bibitem[Henize(1949)]{Henize-1949}
 K.G.\,Henize, \emph{Astrophys. J.}, \textbf{54}, 89 (1949)

\bibitem[Joy(1954)]{Joy-1954}
 A.H.\,Joy, \emph{Astrophys. J.}, \textbf{120}, 377 (1954)

\bibitem[Patterson(1979)]{Patterson-1979}
 J.\,Patterson, \emph{Astrophys. J.}, \textbf{234}, 978 (1979)

\bibitem[Patterson-etal(1980)]{Patterson-etal-1980}
 J.\,Patterson, D.\,Branch, G.\,Chincarini, E.L.\,Robinson, \emph{Astrophys. J.}, \textbf{240}, L133 (1980)

\bibitem[Eracleous-etal(1994)]{Eracleous-etal-1994}
 M.\,Eracleous, K\,Horne, E.L.\,Robinson, et~al., \emph{Astrophys. J.}, \textbf{433}, 313 (1994)

\bibitem[de-Jager-etal(1994)]{de-Jager-etal-1994}
 O.C.\,de~Jager, P.J.\,Meintjes, D.\,O'Donoghue, E.L.\,Robinson, \emph{Monthly. Not. Roy. Astron. Soc.} \textbf{267}, 577 (1994)

\bibitem[Welsh(1999)]{Welsh-1999}
 W.F.\,Welsh, \emph{Proc. Annapolis Workshop on Magnetic Cataclysmic Variables}, eds. C.~Hellier and K.~Mukai, ASP Conference Series \textbf{157}, 357 (1999)

\bibitem[Wynn-etal.(1997)]{Wynn-etal-1997}
 U.A.\,Wynn, A.R.\,King, K.\,Horne, \emph{Monthly. Not. Roy. Astron. Soc.} \textbf{286}, 436 (1997)

\bibitem[Welsh-etal(1998)]{Welsh-etal-1998}
 W.F.\,Welsh, K,\,Horne, R.\,Gomer, \emph{Monthly. Not. Roy. Astron. Soc.} \textbf{298}, 285 (1998)

\bibitem[Reinsch-Beuermann(1994)]{Reinsch-Beuermann-1994}
 K.\,Reinsch, K.\,Beuermann, \emph{Astron. and Astrophys.} \textbf{282}, 493 (1994)

\bibitem[van-Paradijs-etal(1989)]{van-Paradijs-etal-1989}
 J.\,van~Paradijs, H.\,Kraakman, S.\,van~Amerongen, \emph{Astron. and Astrophys. Suppl. Ser.} \textbf{79}, 205 (1989)

\bibitem[Bruch(1991)]{Bruch-1991}
 A.\,Bruch, \emph{Astron. and Astrophys.} \textbf{251}, 59 (1991)

\bibitem[Welsh-etal(1995)]{Welsh-etal-1995}
 W.F.\,Welsh, K.\,Horne, R.\,Gomer, \emph{Monthly. Not. Roy. Astron. Soc.} \textbf{275}, 649 (1995)

\bibitem[de-Jager(1994)]{de-Jager-1994}
  O.C.\,de~Jager, \emph{Astrophys. J. Suppl.}, \textbf{90}, 775 (1994)

\bibitem[Osborne-etal(1995)]{Osborne-etal-1995}
 J.P.\,Osborne, K.L.\,Clayton, D.\,O'Donoghue, et~al., \emph{Proc. Magnetic Cataclysmic Variables}, Eds. D.\,Buckley and B.\,Warner, ASP Conference Series \textbf{85}, 368 (1995)

\bibitem[Choi-etal(1999)]{Choi-etal-1999}
 C.-S.\,Choi, T.\,Dotani, P.C.\,Agrawal, Astrophys. J. {\bf 525}, 399 (1999)

\bibitem[Choi-Dotani(2006)]{Choi-Dotani-2006}
 C.-S.\,Choi, T.\,Dotani, \emph{Astrophys. J.}, \textbf{646}, 1149 (2006)

\bibitem[Itoh-etal(2006)]{Itoh-etal-2006}
 K.\,Itoh, S.\,Okada, M.\,Ishida, H.\,Kunieda, \emph{Astrophys. J.}, \textbf{639}, 397 (2006)

\bibitem[Mauche(2009)]{Mauche-2009}
 C.W.\,Mauche, \emph{Astrophys. J.}, \textbf{706}, 130 (2009)

\bibitem[Bastian-etal(1988)]{Bastian-etal-1988}
 T.S.\,Bastian, G.A.\,Dulk, G.\,Chanmugam, \emph{Astrophys. J.}, \textbf{324}, 431 (1988)

\bibitem[Abada-Simon-etal(2005)]{Abada-Simon-etal-2005}
 M.\,Abada-Simon, J.\,Casares, A.\,Evens, et~al., \emph{Astron. and Astrophys.} \textbf{433}, 1063 (2005)

\bibitem[van-der-Laan(1966)]{van-der-Laan-1966}
 H.\,van der Laan, \emph{Nature}, \textbf{211}, 1131 (1966)

\bibitem[Terada-etal(2008)]{Terada-etal-2008}
 Y.\,Terada, T.\,Hayashi, M.\,Ishida, \emph{Publ. Astron. Soc. Japan}, \textbf{60}, 387 (2008)

\bibitem[Terada(2010)]{Terada-2010}
 Y.\,Terada, T.\,Dotani, \emph{Proc. High energy emission from pulsars and their systems}, eds. N.\,Rea, D.F.\,Torres, Astrophys. and Space Sci. Proc. (Berlin Heidelberg: Springer, 2010), p.\,563

\bibitem[Meintjes-etal(1994)]{Meintjes-etal-1994}
 P.J.\,Meintjes, O.C.\,de\,Jager, B.C.\,Raubenheimer, et~al., \emph{Astrophys. J.}, \textbf{434}, 292 (1994)

\bibitem[Chadwick-etal(1995)]{Chadwick-etal-1995}
 P.M.\,Chadwick, J.E.\,Dickinson, M.R.\,Dickinson, et~al., \emph{Astroparticle Physics}, \textbf{4}, 99 (1995)

\bibitem[Lang-etal(1998)]{Lang-etal-1998}
 M.J.\,Lang, J.H.\,Buckley, D.A.\,Carter-Lewis, et~al., \emph{Astroparticle Physics}, \textbf{9}, 203 (1998)

\bibitem[Ikhsanov-Biermann(2006)]{Ikhsanov-Biermann-2006}
 N.R.\,Ikhsanov, P.L.\,Biermann,  \emph{Astron. and Astrophys.} \textbf{445}, 305 (2006)

\bibitem[Sidro-etal(2008)]{Sidro-etal-2008}
 N.\,Sidro, J.\,Cortina, C.W.\,Mauche, et~al., \emph{Proc. 30th International Cosmic Ray Conference}, eds.  R.\,Caballero, et~al., (Mexico, 2008), \textbf{2}, p.\,715

\bibitem[Eracleous-Horne(1996)]{Eracleous-Horne-1996}
 M.\,Eracleous, K.\,Horne, \emph{Astrophys. J.}, \textbf{471}, 427 (1996)

\bibitem[Beskrovnaya-etal(1996)]{Beskrovnaya-etal-1996}
 N.G.\,Beskrovnaya, N.R.\,Ikhsanov, A.\,Bruch, N.M.\,Shakhovskoy,  \emph{Astron. and Astrophys.} \textbf{307}, 840 (1996)

\bibitem[Abada-Simone-etal(1995)]{Abada-Simone-etal-1995}
 M.\,Abada-Simon, T.S.\,Bastian, K.\,Horne, \emph{Proc. Magnetic Cataclysmic Variables}, eds. D.~Buckley and B.~Warner, ASP Conference Series, \textbf{85}, 355 (1995)

\bibitem[Lamb-Patterson(1983)]{Lamb-Patterson-1983}
 D.Q.\,Lamb, J.\,Patterson, \emph{Proc. Cataclysmic variables and related objects}, Seventy-second Colloquium, Haifa, 1982 (Dordrecht: D. Reidel Publishing Co., 1983), p.\,229

\bibitem[Cropper(1986)]{Cropper-1986}
 M.\,Cropper,  \emph{Monthly. Not. Roy. Astron. Soc.}, \textbf{222}, 225 (1986)

\bibitem[Chanmugam-Frank(1987)]{Chanmugam-Frank-1987}
 G.\,Chanmugam, J.\,Frank, \emph{Astrophys. J.}, \textbf{320}, 746 (1987)

\bibitem[Ikhsanov-etal(2004a)]{Ikhsanov-etal-2004a}
 N.R.\,Ikhsanov, V.V.\,Neustroev, N.G.\,Beskrovnaya, Astron. and Astrophys. {\bf 421}, 1131 (2004)

\bibitem[Ikhsanov-etal(2004b)]{Ikhsanov-etal-2004b}
N.R.\,Ikhsanov, V.V.\,Neustroev, N.G.\,Beskrovnaya, \emph{Astronomy Letters}, \textbf{30}, 675 (2004)

\bibitem[Hameury-etal(1986)]{Hameury-etal-1986}
 J.-M.\,Hameury, A.R.\,King, J.-P.\,Lasota, \emph{Monthly. Not. Roy. Astron. Soc.}, \textbf{218}, 695 (1986)

\bibitem[Ikhsanov(2006)]{Ikhsanov-2006}
 N.R.\,Ikhsanov, \emph{Astrophys. J.}, \textbf{640}, L59 (2006)

\bibitem[Shvartsman(1970)]{Shvartsman-1970}
V.F.\,Shvartsman, \emph{Radiophysics and Quantum Electronics}, \textbf{13}, 1428 (1970)

\bibitem[Lipunov(1980)]{Lipunov-1980}
V.M.\,Lipunov, \emph{Soviet Astronomy} \textbf{24}, 722 (1980)

\bibitem[Davies-Pringle(1981)]{Davies-Pringle-1981}
 R.E.\,Davies, J.E.\,Pringle, \emph{Monthly. Not. Roy. Astron. Soc.}, \textbf{196}, 209 (1981)

\bibitem[Pearson-etal(2003)]{Pearson-etal-2003}
 K.J.\,Pearson, K.\,Horne, W.\,Skidmore, \emph{Monthly. Not. Roy. Astron. Soc.}, \textbf{338}, 1067 (2003)

\bibitem[Ikhsanov-Beskrovnaya(2008)]{Ikhsanov-Beskrovnaya-2008}
 N.R.\,Ikhsanov, N.G.\,Beskrovnaya, \emph{e-Print} arXiv:0809.1169 (2008)

\bibitem[Gosling-etal(1991)]{Gosling-etal-1991}
  J.T.\,Gosling, M.F.\,Thomsen, S.J.\,Bame, et~al., \emph{J. Geophys. Res.} \textbf{96}, 14097 (1991)

\bibitem[Lipunov(1987)]{Lipunov-1987}
   Lipunov, V.M. 1992, \emph{Astrophysics of neutron stars}, Springer-Verlag, Heidelberg

\bibitem[Usov(1988)]{Usov-1988}
  V.V.\,Usov, \emph{Soviet Astronomy Letters}, \textbf{14}, 258 (1988)

\bibitem[Landau-Lifshits(1973)]{Landau-Lifshits-1973}
 L.D.\,Landau, E.M.\,Livshitz, \emph{The Classical Theory of Fields}, \textbf{2}, 4th ed. Butterworth-Heinemann, (1975)

\bibitem[Ikhsanov(1998)]{Ikhsanov-1998}
 N.R.\,Ikhsanov, \emph{Astron. and Astrophys.} \textbf{338}, 521 (1998)

\bibitem[Ikhsanov-etal(2002)]{Ikhsanov-etal-2002}
 N.R.\,Ikhsanov, S.\,Jordan, N.G.\,Beskrovnaya, N.G. \emph{Astron. and Astrophys.} \textbf{385}, 152 (2002)

\bibitem[Jordan(1992)]{Jordan-1992}
 S.\,Jordan, \emph{Astron. and Astrophys.} \textbf{265}, 570 (1992)

\bibitem[Artymovich-Sagdeev(1979)]{Artymovich-Sagdeev-1979}
L.A.\,Artsimovich, R.Z.\,Sagdeev, \emph{Plasma physics for physisists} (M. Atomizdat, 1979)

\bibitem[Arons-Scharlemann(1979)]{Arons-Scharlemann-1979}
 J.\,Arons, E.T.\,Scharlemann, \emph{Astrophys. J.}, \textbf{231}, 854 (1979)

\bibitem[Schoenberner-etal(2000)]{Schoenberner-etal-2000}
 D.\,Sch\"onberner, T.\,Driebe, T.\,Bl\"ocker, \emph{Astron. and Astrophys.} \textbf{356}, 929 (2000)

\bibitem[Livio(1995)]{Livio-1995}
 M.\,Livio, \emph{Proc. Millisecond Pulsars. A Decade of Surprise, Colorado}, 1994, eds. A.S.\,Fruchter, M.\,Tavani, D.C.\,Backer (San Francisco: Astronomical Society of the Pacific Publisher, 1995)

\bibitem[Livio-Pringle(1998)]{Livio-Pringle-1998}
 M.\,Livio, J.E.\,Pringle, \emph{Astrophys. J.}, \textbf{505}, 339 (1998)

\bibitem[Meintjes(2002)]{Meintjes-2002}
 P.J.\,Meintjes, \emph{Monthly. Not. Roy. Astron. Soc.}, \textbf{336}, 265 (2002)

\bibitem[Ikhsanov(1999)]{Ikhsanov-1999}
 N.R.\,Ikhsanov, \emph{Astron. and Astrophys.} \textbf{347}, 915 (1999)

\bibitem[Bisnovatyi-Kogan-Komberg(1974)]{Bisnovatyi-Kogan-Komberg-1974}
G.S.\,Bisnovatyi-Kogan, B.V,\,Komberg, \emph{Soviet Astronomy} \textbf{18}, 217 (1974)

\bibitem[Wickramasinghe(2006)]{Wickramasinghe-2006}
  D.T.\,Wickramasinghe, Private communication (2006)

\bibitem[Konar-Choudhuri(2004)]{Konar-Choudhuri-2004}
 S.\,Konar, A.R.\,Choudhuri, \emph{Monthly. Not. Roy. Astron. Soc.}, \textbf{348}, 661 (2004)

\bibitem[Lovelace-etal(2005)]{Lovelace-etal-2005}
 R.V.E.\,Lovelace, M.M.\,Romanova, G.S.\,Bisnovatyi-Kogan, \emph{Astrophys. J.}, \textbf{625}, 957 (2005)

\bibitem[Cumming(2002)]{Cumming-2002}
 A.\,Cumming, \emph{Monthly. Not. Roy. Astron. Soc.}, \textbf{333}, 589 (2002)

\bibitem[Bisnovatyi-Kogan(2006)]{Bisnovatyi-Kogan-2006}
G.S.\,Bisnovatyi-Kogan,  \emph{Physics Uspekhi}, \textbf{49} (1), 53 (2006)

\bibitem[Yakovlev-Urpin(1980)]{Yakovlev-Urpin-1980}
D.G.\,Yakovlev, V.A.\,Urpin, \emph{Soviet Astronomy} \textbf{24},  126 (1980)

\bibitem[Potekhin-etal(1995)]{Potekhin-etal-1995}
 A.Y.\,Potekhin, D.A.\,Baiko, P.\,Haensel, D.G.\,Yakovlev, \emph{Astron. Astrophys.}, \textbf{346}, 345 (1995)

\bibitem[Andronov-Yavorskij(1990)]{Andronov-Yavorskij-1990}
 I.L.\,Andronov, Yu.B.\,Yavorskij, \emph{Contr. Astron. Obs. Skalnate Pleso}, \textbf{20}, 155 (1990)

\bibitem[Ikhsanov(1997)]{Ikhsanov-1997}
 N.R.\,Ikhsanov, \emph{Astron. Astrophys.}, \textbf{325}, 1045 (1997)

\end{thebibliography}
\end{document}